\documentclass[11pt,twoside]{article}

\usepackage{asp2006}

\usepackage{graphics,graphicx}

\begin{document}

\title{Evolution of a Powerful Radio Loud Quasar 3C~186 and
its Impact on the Cluster Environment at z=1}   %%% Fill in title

\author{Aneta Siemiginowska, Thomas L. Aldcroft, Doug Burke}

\affil{Center for Astrophysics, 
60 Garden St., Cambridge, MA 02138}    %%% Fill in author affiliations

\author{Jill Bechtold} 
\affil{Steward Observatory, University of Arizona, Tucson, AZ}

\author{C.C.Cheung\altaffilmark{1}}

\affil{KIPAC, Stanford University, Stanford CA 94305}

\author{Stephanie LaMassa}
\affil{Johns Hopkins University, Baltimore, MD}

\author{Diana M. Worrall}   %%% Fill in author names
\affil{Department of Physics, University of Bristol, Tyndall Ave., Bristol, UK}

\altaffiltext{1}{Jansky Postdoctoral Fellow of the National Radio
Astronomy Observatory}

\begin{abstract} %%% Abstract to run on from here.

X-ray cluster emission has been observed mainly in clusters with
``inactive" cD galaxies ($L_{bol} \sim 10^{40}-10^{43}$
erg~sec$^{-1}$), which do not show signs of accretion onto a SMBH.
Our recent {\it Chandra} discovery of $>$100~kpc scale diffuse X-ray
emission revealed the presence of an X-ray cluster associated with the
radio loud quasar 3C~186 at redshift z=1.1 and suggests interactions
between the quasar and the cluster. In contrast to the majority of
X-ray clusters the 3C~186 cluster contains a quasar in the center
whose radiative power alone exceeds that which would be needed to
quench the cluster cooling. We present the {\it Chandra} X-ray data and new
deep radio and optical images of this cluster. The 3C~186 quasar is a
powerful Compact Steep Spectrum radio source expanding into the
cluster medium. The 2~arcsec radio jet is unresolved in the {\it Chandra}
observation, but its direction is orthogonal to the elliptical surface
brightness of the cluster. The radio data show the possible presence
of old radio lobes on 10 arcsec scale in the direction of the radio
jet. We discuss the nature of this source in the context of
intermittent radio activity and the interaction of the young expanding
radio source with the cluster medium.
\end{abstract}

%%% MAIN BODY OF TEXT GOES HERE. CONSULT "INSTRUCTIONS FOR AUTHORS USING

%%% LATEX2E MARKUP", SECTIONS 2.3-2.6 FOR HELP WITH EQUATIONS, FIGURES,

%%% AND TABLES.

\section{Introduction}   %%% Top level section head (remove "%" symbol)

We present an X-ray cluster at redshift z=1.1 associated with the
radio loud quasar 3C~186. Diffuse X-ray emission surrounding a
luminous quasar has only been detected around a handful of
low-redshift ($z<0.4$) radio-loud quasars (\citet{siemi2005},
\citet{stockton2006}) and around a few $z>0.5$ radio galaxies
\citep{belsole}.  3C~186 cluster is at high redshift. Its X-ray
surface brightness is high enough to allow for measuring the cluster
redshift and constraining the properties of the cluster at
$z\sim1$. This paper discusses the properties of the quasar, the
cluster and their interactions.

%%%%%% FIGURE 1 - Chandra Smoothed Image
\begin{figure}
\begin{center}
\includegraphics[width=4.2in]{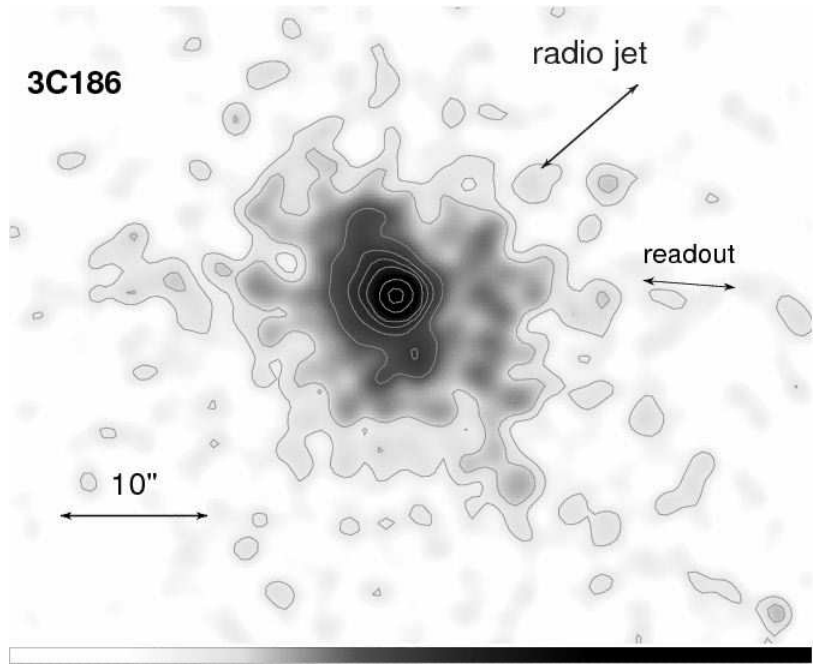}
\caption{Smoothed {\it Chandra} ACIS-S (0.3-7~keV) image of 3C~186
The diffuse cluster emission extends to $\sim$120~kpc from the central
quasar. The 10 arcsec (82~kpc) size is marked.  The direction of an
unresolved 2~arcsec radio jet is also marked; it is orthogonal to the
elliptical distribution of the cluster X-ray emission.}
\end{center}
\label{fig1}
\end{figure}

%%%%%%%%%%% END FIG1

%\subsection{}   %%% Second level section head (remove "%" symbol)

\section{3C186: Quasar}

\begin{figure}
\begin{center}
\includegraphics[width=9.cm]{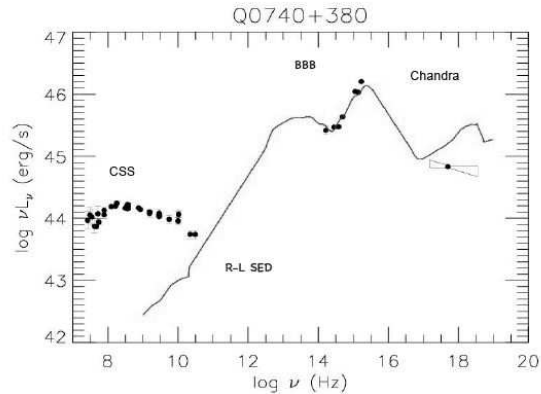}
\caption{Spectral energy distribution of the 3C~186 quasar. 
The {\it Chandra} data are plotted with the 1$\sigma$ bow-tie
regions. The radio and optical photometric data points are taken from
NED. The solid line represents the average radio-loud quasar SED from
\citet{elvis} normalized at log $\nu = 14.5$.
}
\end{center}
\label{fig2}
\end{figure}

3C~186 is a very luminous quasar (L$_{bol} \sim 10^{47}$
erg~sec$^{-1}$). It has a strong big blue bump (BBB) in the optical-UV
band and broad optical emission lines.  The quasar SED is shown in
Fig.~\ref{fig2} The strong big-blue bump emission is generally
interpreted as thermal emission from an accretion disk surrounding the
SMBH. The total luminosity of the 3C~186 BBB is equal to $5.7 \times
10^{46}$~erg~s$^{-1}$. Assuming accretion at the Eddington limit we
obtain a black hole mass of 4.5$\times 10^8$M$_{\odot}$. Using the
measurement of CIV FWHM from
\citet{kuraszkiewicz2002} and the scaling from reverberation mapping
\citep{vester2002} the estimated mass of black hole is larger and equal to
3.2$\times 10^9$M$_{\odot}$.  Thus the central SMBH is large and given
the required accretion rate of $\sim 10$M$_{\odot}$~yr$^{-1}$ to power
this system it is still growing.

%%%%%%%% FIGURE 2

%%%%%%%%%%% END FIG2

%%%%%%%%%% Figure 2 - SED

%%%%%%%%%%

The quasar is extremely radio loud (radio loudness equal to ${\rm Log}
(F_{5{ \rm GHz}}/M_B) = 4.3$) as shown in Fig.\ref{fig2} where the
average radio-loud quasars SED from \citet{elvis} is drawn for a
comparison.  The radio morphology in Fig.~3 shows two components
separated by 2$\arcsec$ and a jet connecting the core and NW
component.
\citet{murgia} estimated the age of the source to be of the
order of $\sim 10^5$~years based on the spectral age of the radio
source.

%%%%%%%% FIGURE 3 - Radio
\begin{figure}[t]
\begin{center}
\includegraphics[width=4.4in]{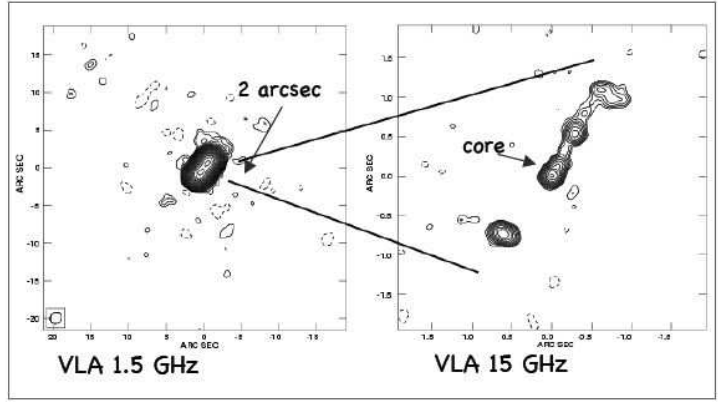}
\vspace*{0.1in}
\caption{{\bf Left:} VLA 1.5 GHz image of 3C~186 from
reprocessing of the archival datasets published in van Breugel et
al. (1992). The restoring beam is 1.62'' $\times$ 1.44'' at position
angle --42.7 degrees shown at bottom left. The image peak is
565~mJy/bm and contour levels begin at 0.5 mJy/beam (2$\sigma$) and
increase by factors of $\sqrt{2}$. Some extended radio emission is
apparent, though not at the angular scale of the observed extended
X-rays. North is up East is left.  {\bf Right:} High resolution
(0.15$\arcsec$) VLA 15~GHz image of 3C~186 showing two lobes and a jet
connecting the core with the Northern Lobe. This structure is entirely
located within a 2 arcsec ($\sim$16~kpc) region.  The image peak is 21.6
mJy/beam, and contours begin at 0.65 mJy/beam increasing by factors of
$\sqrt{2}$.}
\end{center}
\label{radio}
\end{figure}

%%%%%%%%%%%%

\section{3C186: Cluster}

\subsection{X-ray Cluster}

{\it Chandra} detected $\sim$740 source counts in a diffuse emission
surrounding the quasar. We fitted the X-ray spectrum and determined a
cluster redshift of $z=1.1\pm 0.1$ and estimated the X-ray cluster
parameters (see \citet{siemi2005}).  The temperature of
5.2$^{+1.3}_{-0.9}$~keV and gas mass fraction of $\sim 10\%$ are
typical of other clusters at this high redshift
\citep{vikhlinin2002}.
Modeling of the cluster surface brightness gives a normal value of
$\beta$=0.64$^{+0.11}_{-0.07}$, but a relatively small core radius of
$\sim$45$\pm12$~kpc.  The cluster luminosity, $L_{(0.5-2 keV)} = 6
\times10^{44}$~erg~sec$^{-1}$ is typical for the measured cluster
temperature.

Diffuse X-ray cluster emission is detected out to $\sim 15$~arcsec
($\sim$ 120~kpc, Fig.~\ref{fig1}).  The X-ray cluster emission is
non-symmetric and elongated in the NE-SW direction.  This structure is
orthogonal to the 2~arcsec radio jet unresolved in the {\it Chandra}
observation.  The hardness ratios indicate possible spectral
differences within the NE-SW sectors with the harder emission towards
the NE.  Fitting a thermal model shows temperature variations between
3.7~keV and 4.9~keV, but with large errors $\pm$1.5-2~keV.

%%%%%%%%%% FIgure 4 - Optical Image

\begin{figure}[t]
\begin{center}
\includegraphics[width=7.cm]{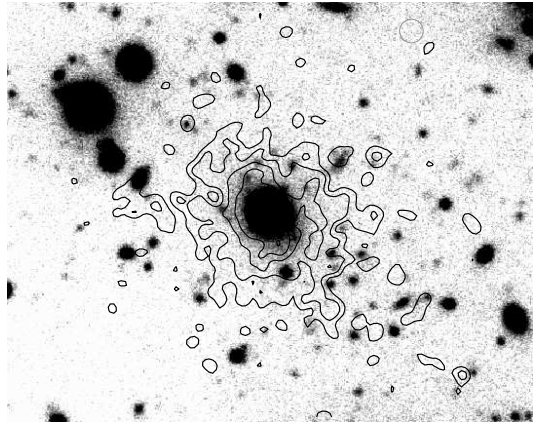}
\caption{\footnotesize
{The {\it Chandra} contours of the X-ray cluster emission superimposed
on the Gemini-GMOS r'-band image.  The bright object in the center is
the quasar, and numerous faint candidate cluster members are visible
around it.}}
\end{center}
\label{fig4}
\end{figure}

%%%%%%%%%%

\subsection{Optical Cluster}

Quasars are usually observed to be offset from the optical peak
emission in rich clusters at high redshift \citep{barr2003}
suggesting loosely bound systems such as groups or proto-clusters, in
the process of forming and undergoing mergers. The optical image
(GEMINI, GN=2007A-Q-110) of the cluster taken in February 2007 is
overlayed with the {\it Chandra} X-ray contours and is presented in
Fig.4. There are many faint candidate cluster members in the field and
their distribution will provide information about the structures
formed in this cluster. We currently have a program to measure
redshifts of the galaxies in the cluster field.

\section{Quasar-Cluster Interaction}

The 3C~186 radio jet and two hot spots are not resolved in the {\it
Chandra} X-ray image. The hot spots are separated by 16~kpc and
located outside the host galaxy. Thus the jet is moving within the
cluster medium. There are two aspects of the possible quasar cluster
interaction: (1) cluster impact on the jet motion and (2) a
transfer of the jet power to the cluster medium. The presence of the
cluster allows us also to look for the signature of past quasar activity
in the form of a relic radio source.

\subsection{Jet Progress}

Based on the cluster central density and temperature, we estimate a
thermal pressure of $\sim 5 \times 10^{-11}$~dyn cm$^{-2}$. If this
pressure is higher than the pressure within the expanding radio
components then the cluster gas may be responsible for confining the
radio source and its small size.  Taking the radio spectral index to
be 1, and approximating each component as a homogeneous spheroid, we
estimate the minimum pressure in each radio component to be
$\sim$10$^{-8}$ dyn cm$^{-2}$.  Thus the radio source is highly
overpressured by about 2-3 orders of magnitude with respect to the
thermal cluster medium and its expansion has not been affected by the
medium.

The jet might also interact and be stopped by a clumpy cold medium of
the host galaxy (Carvalho et al 1998, De Young 1991) and clouds with
densities of 1-30~cm$^{-3}$ are required to confine a jet.  To
estimate the total size of the clumpy medium we can use the limit to
the total X-ray absorbing column density intrinsic to the quasar. The
X-ray absorption limit of $N_H < 9 \times 10^{20}$ cm$^{-2}$ gives the
size of any clumpy medium along the line of sight of order 10-100~pc
compared with the 16~kpc diameter of the radio source. Any such region
cannot significantly limit the expansion of the radio source (see also
discussion in
\citet{matteo}).

\subsection{Cluster Heating}

The radio components are overpressured with respect to the thermal
cluster gas. Thus the expansion of these components into the cluster
medium could potentially heat the center of this cluster. The energy
dissipated into the cluster by the expanding radio components has been
widely discussed in the context of low redshift clusters, where
there is evidence for repetitive outbursts of an AGN. However, the
details of the dissipation process are undecided.

We estimate the energy content of the hot cluster gas assuming a total
emitting volume of 2.3$\times 10^{71}$cm$^3$ and $kT \sim 5$~keV, to
be of the order of ${3\over2} kTnV \sim$ 2$\times 10^{61}$ ergs.
Using the 151~MHz flux
density of 5.9$\times 10^{-24}$ erg~sec$^{-1}$~cm$^{-2}$~Hz$^{-1}$
\citet{hales1993}\citet{willott1999} scaling
which accounts for the total radio emission from the jet and hot
spots, the jet kinetic power is of order $L_{jet} \sim
10^{46}$erg~sec$^{-1}$. If the expanding radio source dissipated this
jet energy
into the cluster's central 120~kpc region, then the heating time would
be $\sim$10$^8$ years.  We can also estimate the amount of mechanical
work done by the jet and radio components during the expansion to the
current radio size ($2\arcsec
\times 0.3\arcsec \sim 2.3 \times 10^{66}$cm$^3$) as $pdV \sim 
10^{56}$~ergs.  If the expansion velocity is of the order of
0.1$c$ then the radio source has been expanding for about $5 \times
10^5$~years with an average power of 6$\times
10^{42}$~erg~sec$^{-1}$. The estimated jet power is $\sim 3$ orders of
magnitude higher.

%%%%%%%%%% FIgure 5 - Radio large scale

\begin{figure}[t]
\begin{center}
\includegraphics[width=4in]{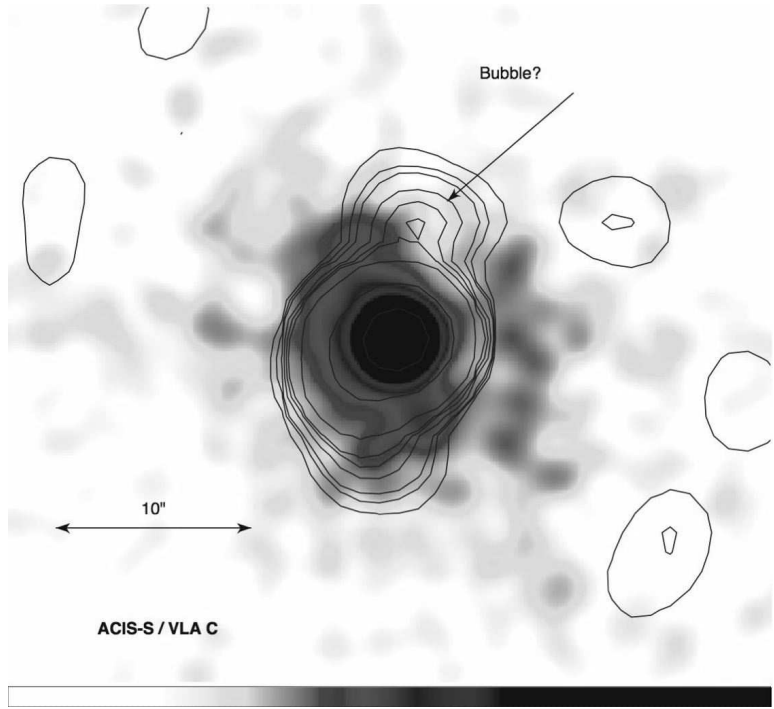}
\caption{\footnotesize
{VLA contours (1.4~GHz, C-configuration, July 2005) overlayed on the
smoothed ACIS-S image of 3C~186. The radio image is ``super-resolved''
 with a 3 arcse beam.
The radio contours start at
0.35 (this is mJy/beam - corresponds to 3*rms) and
continue at 0.7,1,2,3,4,10, 100, 500.
}}
\label{fig5}
\end{center}
\vspace*{-2em}
\end{figure}

\subsection{Relic?}

Large scale radio emission can indicate quasar radio activity in the
past.  We obtained a deep VLA image of 3C~186 in July 2005 to look for
any emission associated with the discovered X-ray cluster. In
Fig.~\ref{fig5} we show the smoothed ACIS-S image overlayed with
1.4~GHz radio contours from our observation. Extended emission is
visible in N-S direction in the radio contours. Note that the
elongation is slightly different than the elongation of X-ray
emission, so the northern radio ``lobe'' points to a location of lower
surface brightness in the cluster.

We estimated a negative signal (``bubble'') in the X-ray data at the
location of the northern radio lobe by comparing a number of counts in
a circular region ($r$=2.5'') centered on this lobe to the background
counts in an annulus at the same distance from the quasar.  We detect
a deficit of -13.9 counts (1.5$\sigma$) at that location.

\section{Summary}

\begin{itemize}

\item We detect a powerful radio loud quasar in an X-ray cluster at redshift $z=1.1$. 

\item Based on radio and X-ray observations we conclude that the radio source has not 
been confined by the cluster gas.

\item The source is at an early stage of expansion.

\item Possible relic activity reflected in the X-ray and radio emission 
will be confirmed with the future {\it Chandra} and VLA
observations.

\item Optical spectroscopy is needed to determine redshifts of the galaxies and the cluster members.

\end{itemize}

\acknowledgements %%% Text of acknowledgements runs on after this command.

This research is funded in part by NASA contract NAS8-39073 and the
{\it Chandra} grants GO-01164X, GO2-3148A and GO-6113X.  The VLA is a
facility of the National Radio Astronomy Observatory is operated by
Associated Universities, Inc. under a cooperative agreement with the
National Science Foundation.

\end{document}